\documentclass{article}

\usepackage{arxiv}

\usepackage[utf8]{inputenc} 
\usepackage[T1]{fontenc}    
\usepackage{hyperref}       
\usepackage{url}            
\usepackage{booktabs}       
\usepackage{amsfonts}       
\usepackage{nicefrac}       
\usepackage{microtype}      
\usepackage{lipsum}		
\usepackage{graphicx}
\usepackage[numbers]{natbib}
\usepackage{doi}
\usepackage{subcaption} 
\usepackage{color} 
\usepackage{adjustbox}
\usepackage{amsmath}
\usepackage{mathtools, nccmath}
\usepackage{xparse}
\usepackage{gensymb} 

\usepackage{multirow}
\usepackage{xcolor}
\usepackage{color,soul}
\usepackage{algorithm}
\usepackage{algorithmic}
\usepackage[normalem]{ulem} 
\usepackage{bm}
\usepackage{bbm}
\usepackage{amssymb}
\usepackage{accents}
\newcommand*{\dt}[1]{%
  \accentset{\mbox{\large\bfseries .}}{#1}}

\title{Definition of dose rate for FLASH pencil-beam scanning
proton therapy: A comparative study}

\usepackage{authblk}

\author[1]{Sylvain Deffet}
\author[2]{Valentin Hamaide}
\author[1,3]{Edmond Sterpin}

\affil[1]{Molecular Imaging, Radiotherapy and Oncology, Institut de Recherche Expérimentale et Clinique (IREC), Université catholique de Louvain, 1200 Woluwe-Saint-Lambert, Belgium}
\affil[2]{Ion Beam Applications SA, Louvain-La-Neuve 1348, Belgium}
\affil[3]{Particle Therapy Interuniversity Center Leuven - PARTICLE, Leuven, 3000, Belgium}

\date{\today}

\renewcommand{\shorttitle}{FLASH dose rate}

\begin{document}
\maketitle

\begin{abstract}
\textbf{Background:} FLASH proton therapy has the potential to reduce side effects of conventional proton therapy by delivering a high dose of radiation in a very short period of time. However, significant progress is needed in the development of FLASH proton therapy. Increasing the dose rate while maintaining dose conformality may involve the use of advanced beam-shaping technologies and specialized equipment such as 3D patient-specific range modulators, to take advantage of the higher transmission efficiency at the highest energy available. The dose rate is an important factor in FLASH proton therapy, but its definition can vary because of the uneven distribution of the dose over time in pencil beam scanning (PBS).\\
\textbf{Purpose:} Highlight the distinctions, both in terms of concept and numerical values, of the various definitions that can be established for the dose rate in PBS proton therapy.\\
\textbf{Methods:} In an \textit{in~silico} study, five definitions of the dose rate, namely the PBS dose rate, the percentile dose rate, the maximum percentile dose rate, the average dose rate, and the dose averaged dose rate (DADR) were analyzed first through theoretical comparison, and then applied to a head and neck case. To carry out this study, a treatment plan utilizing a single energy level and requiring the use of a patient-specific range modulator was employed. The dose rate values were compared both locally and by means of dose rate volume histograms (DRVHs).\\
\textbf{Results:} The PBS dose rate, the percentile dose rate, and the maximum percentile dose are definitions that are specifically designed to take into account the time structure of the delivery of a PBS treatment plan. Although they may appear similar, our study shows that they can vary locally by up to 10\%. On the other hand, the DADR values were approximately twice as high as those of the PBS, percentile and maximum percentile dose rates, since the DADR disregards the periods when a voxel does not receive any dose. Finally, the average dose rate can be defined in various ways, as discussed in this paper. The average dose rate is found to be lower by a factor of approximately 1/2 than the PBS, percentile and maximum percentile dose rates.\\
\textbf{Conclusions:} We have shown that using different definitions for the dose rate in FLASH proton therapy can lead to variations in calculated values ranging from a few percent to a factor of two. Since the dose rate is a critical parameter in FLASH radiation therapy, it is essential to carefully consider the choice of definition. However, to make an informed decision, additional biological data and models are needed.
\end{abstract}

\keywords{FLASH, Flash Proton Therapy, Dose Rate}

\section{Introduction}
FLASH radiation therapy has gained increasing interest in recent years due to its potential to significantly reduce toxicity to normal tissues while maintaining the same level of tumor control when a high dose of radiation is delivered in a short period of time\cite{Favaudon2014,BUONANNO2019,MontayGruel2019}. While more research is needed to fully understand the potential benefits and limitations of FLASH proton therapy as well as the underlying mechanism\cite{Pratx2019,Spitz2019,LABARBE2020,JIN2020}, there is a growing number of research studies on treatment delivery and treatment planning.

One of the main challenges in developing FLASH proton therapy is the need to significantly increase the dose rate. Most preclinical experiments are conducted using small fields that are typically covered by a passive scattering method. However, the achievable field size for a given dose rate is directly limited by the maximum current which can be output by the system. In contrast, pencil beam scanning (PBS) has the potential to locally achieve high dose rates, thanks to the limited size of the spots that are delivered individually. In this paper, we focus solely on the PBS technique, but some of the conclusions can be applied similarly to passive scattering. It is only necessary to consider that the entire field is then achieved with a single, very wide PBS spot.

To achieve a FLASH dose rate, a number of changes are needed to the equipment and delivery system used in intensity modulated proton therapy (IMPT). In particular, the proton beam must be delivered at a much higher energy than is generally used in IMPT to ensure a high transmission efficiency\cite{JOLLY202071}. In addition, using a treatment plan that involves multiple energies incurs delays required by the system to switch from one energy to the next one. These dead times negatively impact the average dose rate. In direct application of these observations, one of the first approaches studied to deliver a FLASH treatment is to shoot at maximum energy through the patient with so-called transmission beams\cite{Rothwell2022}. Of course, a significant amount of dose would also be delivered distal to the tumor, thereby compromising the superior dosimetric potential of protons.

As opposed to simple shoot-through proton therapy, high-energy conformal FLASH proton therapy, may involve the use of advanced beam-shaping technologies and specialized nozzle designs to confine the dose to target volume, similarly to IMPT. To obtain a conformal treatment plan with a single high energy layer, the range of the proton beam must be adjusted by using a range modulator, which is a device that is placed between the nozzle and the patient\cite{JOLLY202071}. The range modulator passively adjusts the energy of the proton beam, which in turn affects the depth at which the protons will deposit their energy in the patient. Several planning methods have been proposed to optimize such range modulators\cite{Simeonov2022, Liu2022, Zhang2022, Deffet2023}.

Besides the calculation of the range modulator, another challenge in planning is to calculate the spot map, ie. the locations and the weights of the PBS spots, in order to maximize the dose rate. In addition, the scanning pattern of the PBS spots may also be optimized\cite{JOSESANTO2022}.

Both for preclinical experiments and for the development of planning and delivery tools, the dose rate is the key parameter in FLASH. However, as a voxel receives its total dose through a series of contributions unevenly distributed over time several factors must be taken into consideration such as the structure of the pulses, the dose delivered by each PBS spot, and the time-related contribution to the dose seen by each voxel.

Calculated over the duration of a pulse, the dose rate can vary by several orders of magnitude, depending on the beam delivery mode as a cyclotron produces a near-continuous beam delivery while a synchrocyclotron produces a pulsed beam with an instantaneous dose rate significantly higher but with a duty cycle significantly lower\cite{JOLLY202071}. A typical isochronous cyclotron delivers a current of up to 800~nA (IBA, Varian) \cite{jones2015experimental, JOLLY202071}. A typical synchro-cyclotron delivers an average current of at least 130~nA (IBA) or 30~nA (Mevion) in pulses of approximately $5-10~\mathrm{\mu s}$ with a period of 1~kHz (IBA) or 500-750~Hz (Mevion)\cite{JOLLY202071}. In all commercial systems, the bunch length is approximately 2~ns\cite{JOLLY202071}. Within one pulse of the IBA synchro-cyclotron, the dose rate is therefore of the order of $10^5~\mathrm{Gy/s}$. On the contrary, within the delivery of one PBS spot the dose rate is of the order of 500~Gy/s - 2000~Gy/s.

It is often assumed that the time scale related to the FLASH effect is larger than the duration of a pulse so that the dose rate may be considered constant during the delivery of a spot\cite{Folkerts2020, hotoiu2021open, pin2022pencil} ($\sim$1-10ms) hence unchanged between pulsed or continuous delivery\cite{Rodriguez2022}, although this remains to be further demonstrated\cite{Rothwell2022}. Nevertheless, in PBS, the total dose received by a voxel is generally the sum of the contributions of several spots. Therefore, the dose rate may be computed independently for each voxel, taking into account the temporal structure of the contributions to the dose, as proposed by Folkerts~\textit{et~al.}\cite{Folkerts2020} which established the so-called definition of the 'PBS dose rate' which is very similar to the concept of percentile dose rate proposed by others\cite{hotoiu2021open, pin2022pencil}.

In the next sections, we first introduce different definitions of dose rate based on a review of the literature. We also generalize the concept of percentile dose rate. These definitions are then compared and discussed based on a head and neck case. Through this \textit{in~silico} experiment, we show how the definition can impact the calculated dose rate values by several percent.

\section{Materials and Methods}
\subsection{Dose rate definitions}
In PBS, the dose is delivered by spots separately temporally by 'beam-off' times required to move the pencil beam from one position to the next. Depending on the voxel in which we compute the dose rate, the temporal structure of contributions to the dose may vary. Therefore, we establish the following definitions which will be used to define the dose rates:

\begin{itemize}
    \item $d_i(t)$ is the dose received in voxel $i$ at time $t$;
    \item $t_i(d)$ is the time elapsed to obtain dose $d$ at voxel $i$;
    \item $D_{i, s}$ is the total dose contribution of spot $s$ to voxel $i$;
    \item $D_i$ it the total accumulated dose received in voxel $i$: $D_i = \sum_s D_{i, s}$;
    \item $T_s$ is the total irradiation time ('beam-on' time) of spot s.
\end{itemize}

When defining the dose rate, some authors consider that 'beam-off' times, or even the time when a voxel does not receive any dose, should not be taken into account, while others include them in their definitions. In preclinical studies involving passively scattered beams, the dose rate is computed for the whole field as the total dose divided by the total duration of the field irradiation. This definition could be extended to scanned beams. However, in our study, we make the assumption that the distribution of the dose rate within the volume of interest matters. Therefore, we consider that the dose rate must be computed per voxel.

One conservative method is to calculate the dose rate for each voxel individually while accounting for periods of time when it does not receive any dose. This approach gave rise to the PBS dose rate, which is introduced below, along with two of its variants.

\subsubsection{PBS dose rate}
The PBS dose rate was defined by Folkerts~\textit{et~al.}\cite{Folkerts2020, Varian2023} to account for the time structure of the delivery of a PBS treatment plan:
\begin{equation}
    \dt{D}^{PBS}_i = \frac{D_i-2D^{Th}}{t_{1, i} - t_{0, i}}
    \label{eq:PBSDR}
\end{equation}
where $t_{1,i}=t_i(D_i-D^{Th})$, $t_{0,i}=t_i(D^{Th})$, and $D^{Th}$ is an arbitrary dose threshold.

This dose threshold is often computed as a percentage of the prescription: $D^{Th} = p \times \mathrm{prescription}$ with $p$ being an arbitrary percentage. For example, to compute the dose rate on an interval corresponding to 95\% of the dose, we would use $p=2.5\%$.

\begin{figure}
    \centering
    \includegraphics[width=0.52\textwidth]{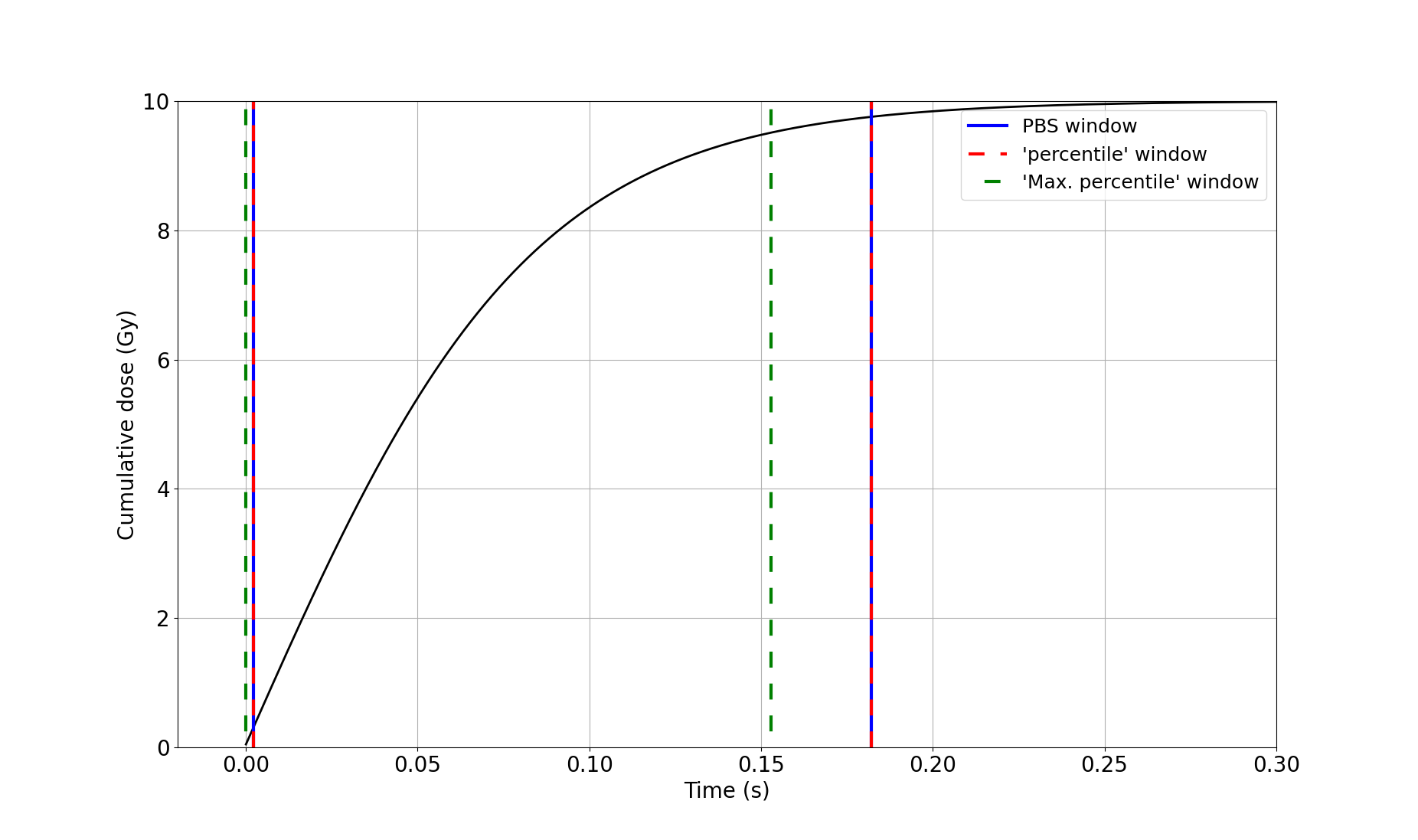}
    \caption{Time windows as used by the PBS, percentile and maximum percentile dose rates for $p=0.025$ and an arbitrary cumulative dose profile corresponding to a prescription of 10~Gy.}
    \label{fig:theor_ex}
\end{figure}

Fig.~\ref{fig:theor_ex} illustrates this definition for $p=0.025$ through a hypothetical example where the total dose received by the voxel is 10~Gy which is also assumed to be the prescription. It can be seen that the time window extends from the moment when the dose reaches 0.25~Gy until the moment when it reaches 9.75 Gy.

The crucial aspect of this definition is the calculation of the time required by the system to deliver a designated dose, for example, 10 Gy, to a voxel. This calculation depends on the voxel's location and incorporates the specific starting time named $t_0$, and ending time named $t_1$, associated with that voxel. Significantly, even if there are periods during the irradiation when the voxel receives a limited or no dose, those intervals are still taken into account since the voxel has not yet received its complete dose. This differs from the concept of dose-averaged dose rate, which will be introduced and discussed later.

\subsubsection{Percentile dose rate}
In the definition of the PBS dose rate, the dose threshold is a fixed value, which can be a percentage of the prescription. This arbitrary choice is questionable outside the target volume where the dose is expected to be much lower than the prescription. In particular, for low doses, the PBS dose rate might be undefined. For example, for a fraction of 10~Gy and $p=0.025$ the PBS dose threshold is 0.25~Gy. Therefore, the PBS dose rate cannot be computed in any voxel presenting a dose lower than two times this value, i.e. 0.5~Gy.

The threshold value may of course be set differently depending on the organ. For example, some consider an individual threshold for each voxel, proportional to the dose delivered to that voxel\cite{hotoiu2021open, pin2022pencil}. In this paper this definition is referred to as the percentile dose rate:
\begin{equation}
    \dt{D}^{PERC}_i = \frac{(1-2p)D_i}{t_{1, i} - t_{0, i}}
    \label{eq:MDR}
\end{equation}
where $t_{1,i}=t_i((1-p)D_i)$, $t_{0,i}=t_i(pD_i)$, and $p$ is an arbitrary percentage.

\subsubsection{Maximum percentile dose rate\label{sec:perc_dr}}
According to both definitions above, the PBS and percentile dose rates are computed over a window of time. The position of the window is defined as the moment, named $t_0$, when the dose exceeds an arbitrary threshold. The definition of time $t_0$ could actually have a significant impact on the calculated value. For example, if we choose $p=2.5\%$, there certainly exist several windows that can cover 95\% of the dose, although this means that they have different starting points and different widths.

The maximum percentile dose rate extends the concept of percentile dose rate by considering all time windows in which the accumulated dose is at least $(1-2p)D_i$:
\begin{eqnarray}
\dt{D}^{MP}_i &=&\max_{t_0, t_1} \frac{\int_{t_0}^{t_1}d_i(t)dt}{t_1 - t_0}\label{eq:ODR}\\
    &s.t.& \int_{t_0}^{t_1}d_i(t)dt \geq (1-2p)D_i \nonumber\\
    && t_1 > t_0 \nonumber
\end{eqnarray}

In this formula, the calculated time window is the one that gives the maximum dose rate while respecting the constraint that the cumulative dose is greater than an arbitrary percentage of the total dose delivered to the voxel.

We can see from the hypothetical example of Fig.~\ref{fig:theor_ex} that this definition provides a different time window than those of the PBS and the percentile dose rates. Indeed, the 95-maximum percentile dose rate window begins at $t=0$, when the cumulative dose is still zero, and ends when the cumulative dose in the voxel reaches 9.5~Gy. Although the cumulative dose is 9.5~Gy for the 3 definitions, the time interval is shorter in the case of the 95-maximum percentile dose rate, which gives a higher dose rate value. This example remains purely theoretical. We will later apply the various definitions to a clinical case in order to discuss the implications of the differences that exist between them.

In our implementation, the maximum percentile dose rate was efficiently computed by means of sliding windows where $t_1$ was defined as $t_1 = t_0 + \mathrm{window~width}$. The sliding of this window over $d_i$ was implemented as a convolution product, for each window width.

At first glance, this implementation might seem similar to the sliding window defined by Schwarz~\textit{et~al.}\cite{Schwarz2022} which proposed a metric related to the 'Flash effectiveness model' proposed by Krieger~\textit{et~al.}\cite{Krieger2022}. However, in the method of Schwarz~\textit{et~al.}, the dose threshold and the dose rate are used as input to define the time window to be slid over the accumulated dose contributions. For example, a threshold of 4~Gy for a dose rate of 40~Gy/s gives a window width of 100~ms. This window would then be slid to determine when the dose is greater than 4~Gy. On the other hand, a threshold of 40~Gy would give a window of 1~s. With this much larger window, dose contributions considered invalid with the 4~Gy threshold could now end up valid since only the dose accumulated over the time window matters. This might seem inconsistent but the main issue here is that the dose rate is not an output value but an input value used to compute the window width. 

\subsubsection{Average dose rate}
The average dose rate can be defined as the total dose delivered to a given voxel divided by the total time required to deliver all the contributions. This definition is equivalent both to the 100-percentile dose rate and to the PBS dose rate with a dose threshold of $0^+~\mathrm{Gy}$:
\begin{equation}
\dt{D}^{100}_i = \frac{D_i}{t_1 - t_0}\label{eq:DR_average}\\
\end{equation}
where $t_1(\mathbf{S}) = t_i(D_i^-)$ and $t_0(\mathbf{S}) = t_i(0^+)$.

The use of a threshold of $0^+~\mathrm{Gy}$ instead of $0~\mathrm{Gy}$ is the mathematical expression that we do not want to count the time when no dose is given before the first dose contribution to the voxel and also after the last contribution to the voxel.

\subsubsection{Dose-averaged dose rate}
The DADR is one of the first dose rate definitions established specifically for FLASH proton therapy by van~de~Water~\textit{et.~al.}, in 2019\cite{Van_de_Water2019-pk}.

In the DADR, the instantaneous dose rate is first averaged over the spot duration, ie. $\frac{D_{i, s}}{T_s}$ is computed for each spot $s$. Then, these contributions to the dose rate are weighted by the dose contribution of each spot to the voxel:
\begin{equation}
    \dt{D}^{DA}_i =\frac{\sum_s D_{i, s} \frac{D_{i, s}}{T_s}}{D_i}
    \label{eq:DADR}
\end{equation}

Consequently, the DADR neglects not only the 'beam-off' times but also the time during which the voxel does not receive any dose because the system is irradiating a distant position.


\subsection{\textit{In~silico} assessment}
The different dose rate definitions were applied to a FLASH conformal treatment plan calculated with an in-house method. In this paper, a treatment plan was optimized on a head and neck case which is reused in the present publication. The PTV considered in the present study had a prescription of 54~Gy, as presented in Fig.~\ref{fig:dr_maps}. However, dose rates must be computed per fraction. As FLASH treatments will most likely be hypofractionated\cite{Van_de_Water2019-pk}, we considered that the dose per fraction would be around 8~Gy.

The treatment plan, and its associated range modulator, were computed using our in-house research treatment planning system, openTPS\cite{opentps}, for an energy of 226~MeV, a spot size of $(\sigma_x = 4.5~\mathrm{mm}, \sigma_y=5~\mathrm{mm})$, and a spot spacing of 15~mm.

To facilitate the comparison of the different dose rate formulas, we used a simple model where:
\begin{enumerate}
    \item the nozzle output current was considered constant;
    \item the time between 2 spots was proportional to the distance between the spots.
\end{enumerate}

In other words,
\begin{enumerate}
    \item the irradiation time of a spot was the ratio between the charge of the spot and the current;
    \item the time separating 2 spots was the ratio between the distance of the spots and an assumed scanning speed.
\end{enumerate}

The spots were ordered in a conventional serpentine pattern. The current was 500~nA (averaged on a pulse period) at the output of the nozzle and the scanning speed was 8000~mm/s. In this study, we assumed an equal scanning speed in all directions. However, it is worth noting that in practical scenarios, the scanning speed typically varies in the x and y directions. Finally, we considered that 1~MU corresponds to 152,880,000 protons.

The Monte Carlo dose engine used in this study is MCsquare\cite{Souris2016} which takes advantage of mutli-core CPU architectures. MCsquare has the ability to compute the dose independently for each beamlet and store it in a sparse matrix without significant increase of the computation time. This feature was used to compute the dose influence matrix, ie. to determine the contributions of each spot to the dose of voxel $i$. The default beam model was adjusted to incorporate the aforementioned characteristics specific to the FLASH energy. Additionally, the default CT calibration was employed, and two extra calibration points were included for the materials of the range modulator (aluminium) and the aperture (tungsten). To avoid resampling artefacts of the range modulators, all computations were done on CTs resampled on the beams-eye views. The CT resolution was $1\times1\times2~\mathrm{mm^3}$. Calculated dose maps were then resampled back on the original CT.

\section{Results}
\begin{figure}
    \centering
    \includegraphics[width=0.45\textwidth]{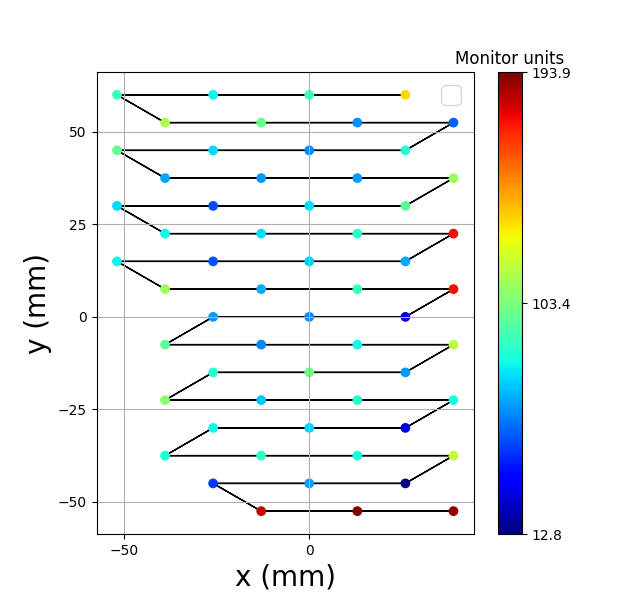}
    \caption{Delivery pattern of the PBS spots}
    \label{fig:spotPattern}
\end{figure}

As emphasized in the previous section, the delivery pattern and time structure of the PBS spots are of utter importance when defining the dose rate, in proton therapy. Factors such as the sequence in which the spots are delivered, the spacing between the spots, the overall shape of the pattern, and the scanning frequency should be considered. Fig.~\ref{fig:spotPattern} shows the scanning pattern and the intensities of the spots. The irradiation time of each spot is proportional to the charge, as a result of our simplified model which relies on the assumption of a constant current.

\begin{figure*}
    \centering
    \includegraphics[width=\textwidth]{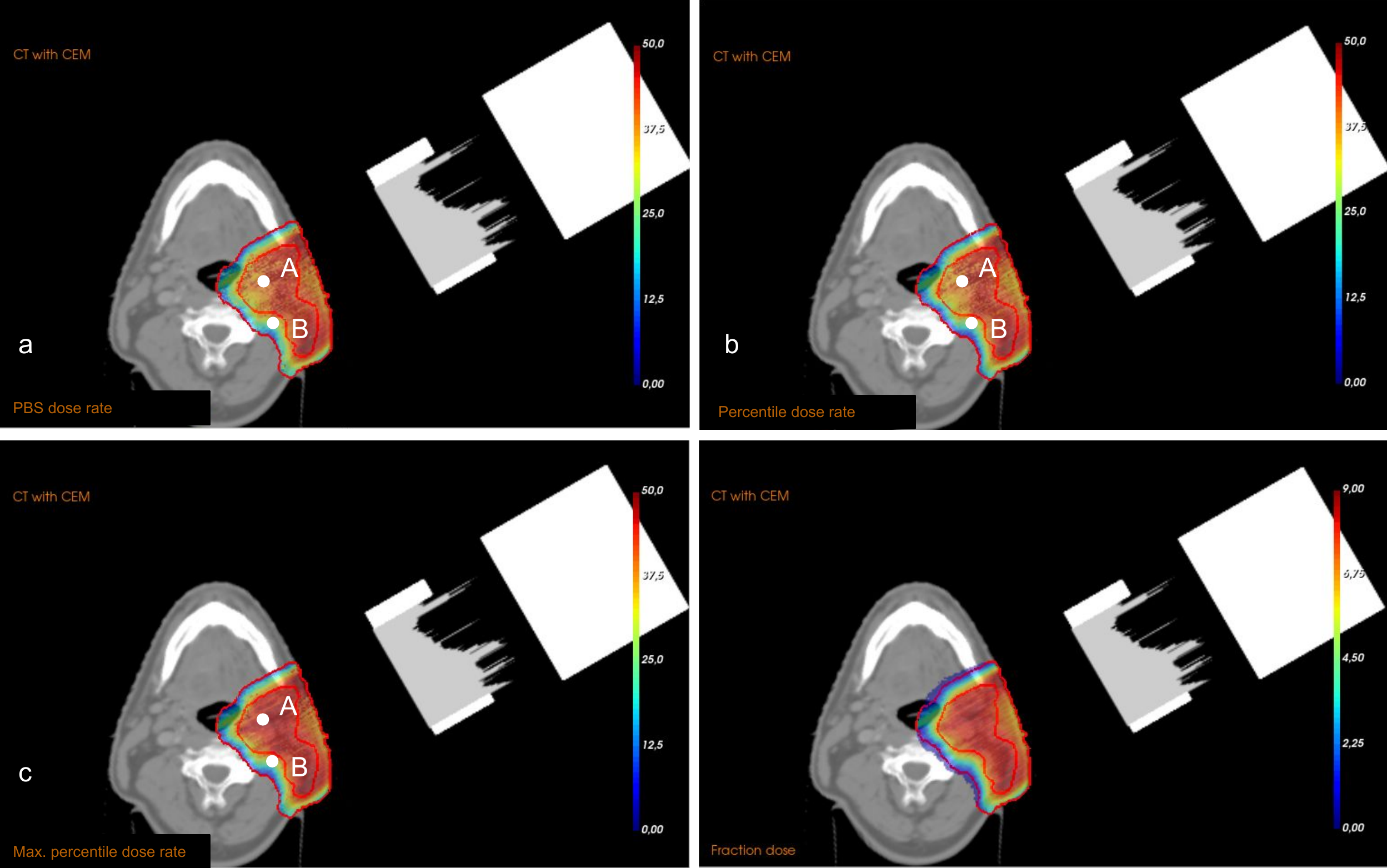}
    \caption{(a) PBS dose rate, (b) 95-percentile dose rate, (c) 95-maximum percentile dose rate, and (d) dose map.}
    \label{fig:dr_maps}
\end{figure*}

The PBS, percentile, and maximum percentile dose rates are shown in Fig.~\ref{fig:dr_maps}a-c. Furthermore, Fig.~\ref{fig:dr_maps}d illustrates the dose corresponding to the fraction of 8 Gy utilized in this study. The white square in the figure corresponds to the range shifter, which is relatively thick given the shallow target volume. It is important to note that the range shifter used in this configuration was made out of aluminium. Positioned between the range shifter and the patient is the patient-specific range modulator, which is enclosed by an aperture. The dose rates shown in the figure were computed using a percentile value of $p=0.025$. As explained previously, this value means that the dose rate was computed over 95\% of the dose. We also evaluated the dose rate for other values ranging from the $p=0.05$ (90-maximum percentile) to $p=0.005$ (99-maximum percentile) and we always observed very similar trends, hence we only show results for $p=0.025$. 

Globally, the maximum percentile dose rate is higher than the PBS and percentile dose rates which look quite similar although differences are noticeable. We selected two locations on this map to better understand those differences.

\begin{figure*}
    \centering
    \includegraphics[width=\textwidth]{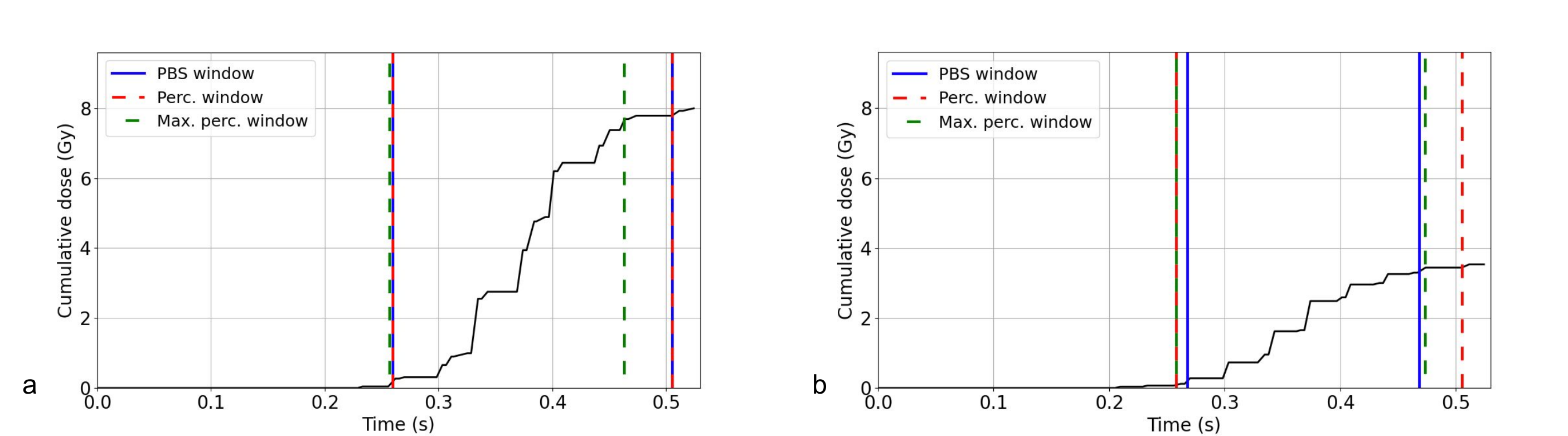}
    \caption{Accumulated dose and time windows as used by the PBS, percentile and maximum percentile dose rates for (a) point~A and (b) point~B defined in Fig.~\ref{fig:dr_maps}.}
    \label{fig:windows}
\end{figure*}

At point A, the percentile dose rate ($30.96~\mathrm{Gy/s}$) is equal to the PBS dose rate but the maximum percentile dose rate ($36.72~\mathrm{Gy/s}$) is higher. In Fig.~\ref{fig:windows}, the plot of the time windows corresponding to each of the dose rates gives a visual explanation. While, by definition, the maximum percentile dose rate selects the shortest window out of all windows that cover 95\% of the dose, for the other two definitions the size of the time window depends on the instants at which the dose exceeds a lower threshold and an upper threshold. In that voxel, the percentile dose rate is equal to the PBS dose rate as the delivered dose perfectly corresponds to the prescription.

At point B however, the PBS dose rate ($16.54~\mathrm{Gy/s}$) is higher as the dose is well below the prescription since this point is outside the PTV. Namely, the dose is 3.5~Gy. However, the PBS dose rate defines the thresholds to be taken into account for the calculation of the time window based on a percentage of the prescription which is 8~Gy. On the contrary, the percentile dose rate ($13.43~\mathrm{Gy/s}$) defines the thresholds based on a percentage of the total dose delivered in the voxel.

As the dose rate varies locally, it seemed natural to present the quantity of volume irradiated under a certain dose rate in the form of a dose rate volume histogram (DRVH), similar to what is done for the dose\cite{Folkerts2020}. It is important to note that, similar to how the DVH alone is not sufficient to fully evaluate the quality of a plan, so is the DRVH with respect to assessing the plan's ability to meet FLASH objectives. Some recent results show that the FLASH effect could depend on both the dose rate and the dose\cite{MONTAYGRUEL2017, Rothwell2022}, and even if a dose threshold exists, it is likely that calculating the dose rate only on voxels in which the dose exceeds this threshold would be too simplistic.

For the evaluation of a FLASH treatment, it seems logical to assess the dose rate in the organs at risk and the healthy tissues surrounding the target volume. However, in order to facilitate the comparison, we did not define a volume for each organ at risk (OAR) but rather a unique volume surrounding the PTV. To draw this volume, we first computed an extension of 15~mm of the PTV (PTV subtracted from the dilated PTV), named PTV-ext, from which we excluded areas where the dose was lower than 1~Gy. Hence, the DRVHs were calculated on two volumes: on the PTV which is supposed to receive a uniform dose of 8~Gy, and on PTV-ext which should receive a lower dose.

\begin{figure*}
    \centering
    \includegraphics[width=\textwidth]{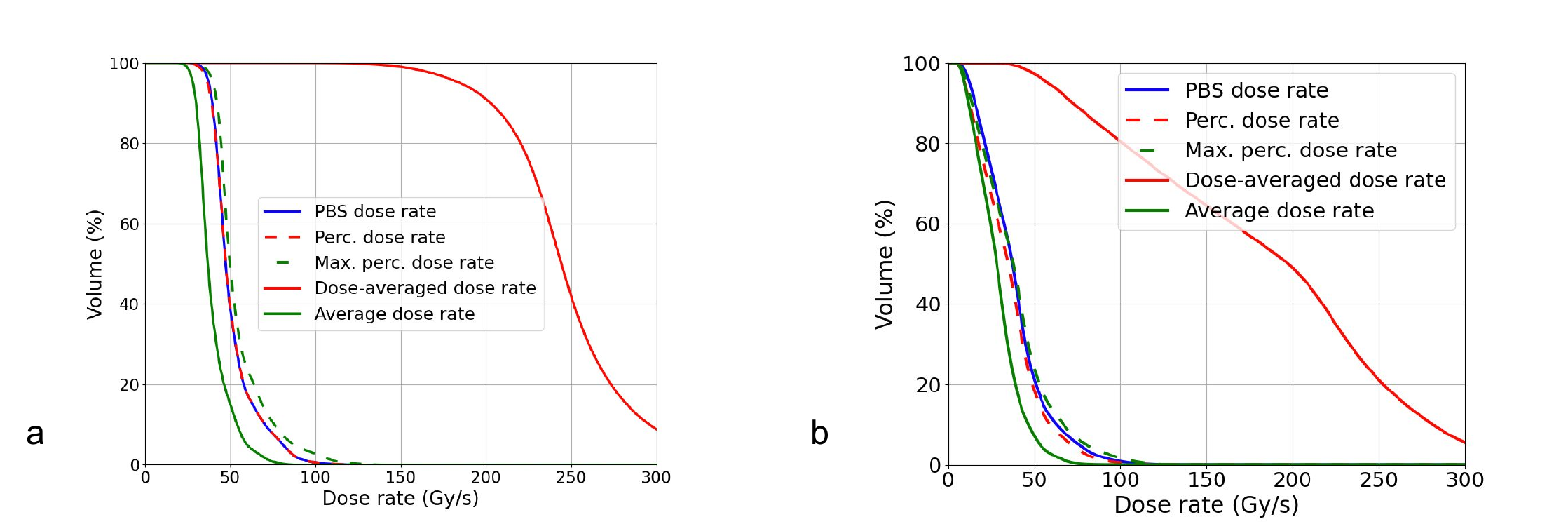}
    \caption{DRVHs computed (a) on the PTV, and (b) on the PTV-ext.}
    \label{fig:drvh}
\end{figure*}

The DRVHs shown in Fig.~\ref{fig:drvh} and the associated metrics listed in Table~\ref{tab:metrtics_perc} confirm the results presented above. Firstly, the PBS and the percentile dose rates are very similar in the PTV since the dose is very close to the prescription. Second, the maximum percentile dose rate is higher than the PBS and percentile dose rates, except in PTV-ext where the the PBS dose rate is higher better in low dose areas which leads to higher DR95 and DR98 values.

While the primary objective of this work is the comparison of the various dose rate definitions, it is worth noting an interesting observation that extends beyond this scope. Specifically, the dose rate in PTV-ext is generally lower compared to the regular PTV. This discrepancy arises from the fact that the dose in the distal region of PTV-ext, which paradoxically corresponds to an area where FLASH could be beneficial, is significantly lower.

\begin{table}
\centering
\begin{tabular}{lrrrrr}
\toprule
{} &   DR2 &   DR5 &  DR50 &  DR95 &  DR98 \\
\midrule
PTV, $(1-2p)=95\%$ \\
\hline
PBS dose rate & 88.38 & 80.95 & 47.24 & 37.39 & 34.96 \\
Percentile dose rate & 88.33 & 81.06 & 47.35 & 36.65 & 33.77 \\
Max. percentile dose rate & 104.36 & 87.32 & 49.76 & 39.88 & 36.81 \\
Av. dose rate & 69.15 & 59.77 & 36.79 & 28.04 & 25.97 \\
DADR & 345.87 & 318.33 & 244.34 & 184.75 & 163.48 \\
\midrule
PTV-ext, $(1-2p)=95\%$ \\
\hline
PBS dose rate & 88.88 & 75.98 & 37.34 & 12.51 & 9.93 \\
Percentile dose rate & 83.28 & 71.66 & 35.03 & 9.69 & 7.68 \\
Max. percentile dose rate & 97.52 & 80.23 & 37.81 & 10.80 & 8.82 \\
Av. dose rate & 62.81 & 53.36 & 28.31 & 9.54 & 7.69 \\
DADR & 329.24 & 302.79 & 197.30 & 58.26 & 46.87 \\
\bottomrule
\end{tabular}
\caption{DRx metrics (Gy/s) of the various definition of dose rates studied in this paper.}
\label{tab:metrtics_perc}
\end{table}

The calculated value of the average dose rate can be significantly affected by the weak contributions at the beginning or end of the irradiation. This is evidenced by the fact that the PBS and the percentile dose rates are several Gy/s higher when these contributions are taken into account.

Finally, as shown on Fig.~\ref{fig:drvh} and in Table~\ref{tab:metrtics_perc}, the DADR values are approximately twice as high as those of the PBS, percentile and maximum percentile dose rates, since the DADR discard, for each voxel, the periods when it does not receive any dose.

\section{Discussion}
In proton therapy, numerous considerations should be taken into account when defining the dose rate in FLASH. One of the most important is the need to take into account the delivery pattern of the PBS spots and to a lesser extent the 'beam-off' times. A voxel in the patient may receive dose contributions at different times. These contributions can be separated by several milliseconds and we showed that the way this is accounted for may impact the value of the computed dose rate by a factor of 2.

The values calculated for the dose rate depend on the treatment plan, including the number, intensity, and order of spots, as well as on the characteristics of the delivery system, such as the current and time between PBS spots. In the current study, we considered a fraction of 8 Gy. For lower doses, the time between PBS spots when the beam is off can have a greater impact on the overall dose rate, while for lower currents, the 'beam-on' times become relatively more important.

The DADR leads to much higher dose rate values as periods which does not bring any contribution to the dose are discarded. As pointed out by others, this lack of consideration of the temporal structure of the dose delivery may provide the same dose rate estimate from an array of spots regardless of the delivery pattern and the time separating the spots\cite{Folkerts2020}. In fact, by strictly conforming to this definition, the switching times between the energy layers of an IMPT plan is not an issue and the use of a range modulator is not necessary, which goes against what seems to be a consensus in the treatment planning community. However, as pointed out by Rothwell~\textit{et.~al.}\cite{Rothwell2022}, the relevant time scale of the dose rate is still unknown as it is intrinsically related to the biology. Given the lack of certainty regarding our understanding of the underlying mechanism of the FLASH effect, we cannot theoretically demonstrate which definition of the dose rate is more suitable. FLASH therapy is still an experimental method under evaluation and more experimental data are needed to answer this question.

This uncertainty should lead us to be cautious about the use of one definition or another. For example, the PBS dose rate, the percentile dose rate, and the maximum percentile dose rate are seemingly very close definitions. However, the corresponding dose rate values may vary locally by several percent. These variations can also be seen in the DRVHs. In our \textit{in~silico} assessment, we had a DR95 of 12.51~Gy/s for the PBS dose rate and of 10.80~Gy/s for the maximum percentile dose rate in PTV-ext whereas, in the PTV, the maximum percentile dose rate was higher than the PBS dose rate. The differences are even more striking when looking at the DR5, where the three definitions lead to very different dose rate values.

Behind these three definitions are two fundamental questions:
\begin{itemize}
    \item Is there a certain amount of dose that can be neglected in the dose rate calculation? If so, can this quantity be assimilated to a fixed value (PBS dose rate) or a percentage of the delivered dose (maximum percentile dose rate, percentile dose rate)?
    \item Does the time window depend on the exact moment at which the dose threshold mentioned above is reached? For example, if the voxels receive 10 Gy and we can ignore 1 Gy, can we use any time window that covers 9 Gy (maximum percentile dose rate)? Or should we use a time window that starts precisely when the dose reaches 0.5 Gy and ends precisely when it reaches 9.5 Gy (PBS dose rate, percentile dose rate)?
\end{itemize}

The first question clearly relies on biological data and models. The second question may seem secondary. However, we have observed that locally the maximum percentile dose rate can be more than 10\% higher than the percentile dose rate evaluated with the same relative threshold. This shows how necessary it is to define exactly what this threshold corresponds to, again through a cross-disciplinary approach.


\section{Conclusions}
Several definitions of the dose rate have been established, which mainly differ in the way that they take into account the spatial and temporal distributions of the PBS spots in proton therapy. We showed that this can lead to differences in the computed dose rate values varying from several percent up to a factor 2. As the dose rate is the key parameter in FLASH, the choice of the definition is very important. However, given the current uncertainty on the time scale related to the FLASH effect, we cannot currently determine which definition is the most suitable from a radiobiological perspective.

\section*{Acknowledgments}
This work was supported by the Walloon Region of Belgium through technology innovation partnership no. 8341 (EPT-1 – Emerging Proton Therapies Phase 1) co-led by MecaTech and BioWin clusters.

\section*{Conflict of Interest Statement}
Dr. Valentin Hamaide was an employee of Ion Beam Applications during the writing of this paper.

\bibliographystyle{unsrt}
\bibliography{biblio}

\end{document}